\documentclass{lhep}       

\def\be{\begin{equation}}
\def\ee{\end{equation}}
\def\bea{\begin{eqnarray}}
\def\eea{\end{eqnarray}}

\def\s{{\mathsf{s}}}

\begin{document}
\title{
Hot vs cold hidden sectors and their effects on thermal relics}
\author{Jinzheng Li,\auno{1} and Pran Nath,\auno{1}}
\address{$^1$ Northeastern University, Boston, Massachusetts 02115-5005, USA}

\begin{abstract}
A variety of possibilities exist for dark matter aside from WIMPS, such as hidden
sector dark matter. We discuss synchronous thermal evolution of visible and hidden sectors and show that the density of thermal relics can change $O(100\%)$ and $\Delta N_{\rm eff}$ by a factor of up to $10^5$ depending of whether the hidden sector was hot  or cold at the reheat temperature. It is also shown that the approximation of using separate entropy conservation for the visible and hidden sectors is invalid even for a very feeble coupling between the two.
\end{abstract}

\maketitle


\section{Introduction}
In exploration of Physics Beyond the Standard Model, hidden sectors play a role
in a variety of settings such as in  supergravity
(for a review see\cite{Nath:2016qzm}), in strings\cite{Candelas:1985en},  in branes\cite{Polchinski:1996na},
and in moose/quiver theories~\cite{Hill:2000mu}.
 Much like the visible sector the hidden sector could contain gauge fields and matter fields and 
 it is altogether possible that dark matter may reside in the hidden sector. 
 The success of the electroweak physics in the standard model indicates that the
 coupling of hidden sector with the visible sector must be feeble. 
On the other hand the coupling of the hidden sector with the inflaton is largely unknown. Thus the couplings of the hidden sector with the inflaton could
be as strong as of the standard model leading to the hidden sector being 
hot with $\xi(T)\equiv \frac{T_h}{T}|_{RH}\simeq 1$, where $T(T_h)$ is the
visible (hidden) sector temperature and $RH$ refers to the reheat temperature
 of the universe.
 Alternately the hidden sector may not couple or may have suppressed couplings with the inflaton in which case   $\xi_0\simeq 0|_{RH}$. It is then of interest to determine the evolution of $\xi(T)=T_h/T$ as a function of $T$. This is of 
 importance since $\xi(T)$ enters in the analysis of observable physics such as the 
   relic density, dark matter cross-sections, $\Delta N_{eff}$ at BBN, and other low energy observables.   
 Recently the evolution equation for $\xi(T)$ has been derived from energy conservation~\cite{Aboubrahim:2020lnr,Aboubrahim:2021ycj,Aboubrahim:2022bzk}, i.e.,  
\bea
\label{rhov}
\frac{d\rho_v}{dt} + 3 H(\rho_v+p_v)&=j_v,~  \text{(visible sector)}\,,\\
\frac{d\rho_h}{dt} + 3 H(\rho_h+p_h)&=j_h,~ \text{(hidden sector)}.
\label{rhoh}
\eea
where $\rho_v (\rho_h)$ is the energy density of the visible (hidden) sector,
$p_v(p_h)$ is the pressure for the visible (hidden sector), $(j_v, j_h)$ are
the sources and $H$ is the Hubble parameter. A straightforward analysis 
leads to the following differential equation for $\xi(T)$
\bea
\frac{d\xi}{dT}= \left[ -\xi \frac{d\rho_h}{dT_h} +
\frac{4H\eta_h\rho_h-j_h}{4H\eta\rho-4H\eta_h\rho_h+ j_h} \frac{d\rho_v}{dT}\right] (T \frac{d\rho_h}{dT_h})^{-1}.
\label{xi1}
\eea
where $\eta=1$ (radiation dominance), $\eta=3/4$ (matter dominance). 
We note in passing that the assumption of separate entropy conservation of 
the visible and the hidden sector to estimate $\xi(T)$ (see, e.g.,\cite{Feng:2008mu,Ertas:2021xeh}) could deviate  substantially from the
true value even for very feeble coupling between the sectors as discussed in 
subsection (\ref{ent1}).\\
 
 \section{ A hidden sector model}
 As a concrete example of a hidden sector, we consider a $U(1)_X$ extension of 
 the standard model  with a particle content consisting of a gauge boson
 ($C_\mu$), a Dirac fermion (D) charged under $U(1)_X$ with a gauge coupling
 constant $g_x$ and spin zero dark fields $\phi, s$. 
 Communication with the visible sector occurs via kinetic
 mixing \cite{Holdom:1985ag} or Stueckelberg mass mixing\cite{Kors:2004dx}
 between the $U(1)_X$ gauge field $C_\mu$ and the  hypercharge $U(1)_Y$ gauge field $B_\mu$ of the standard model. The communication between the two can also take place via a combined kinetic-Stueckelberg-mass mixing\cite{Feldman:2007wj}, via a Stueckelberg-Higgs mixing
  \cite{Du:2022fqv} and via a variety of other mechanisms such as via Higgs
  portal~\cite{Patt:2006fw} and higher dimensional operators.
For the case of kinetic mixing one adds a gauge invariant
combination $\frac{\delta}{2}  C^{\mu\nu} B_{\mu\nu}$, and for the Stueckelberg mass mixing one adds $(m_1 C_\mu + m_2 B_\mu + \partial_\mu \sigma)^2$
where $\sigma$ is an axionic field which transforms dually under $U(1)_X$ and
$U(1)_Y$ to keep the mass mixing term gauge invariant. 
In the mass and kinetic energy diagonal basis for the gauge bosons, 
one will have have a massive dark photon $\gamma'$ with mass $m_{\gamma'}$
in addition to the standard model gauge bosons $W^{\pm}, Z$.
 The mass mixing mechanism generates a milli-charge on the hidden sector matter~\cite{Kors:2004dx,Cheung:2007ut} and such matter is relevant in the explanation of EDGES anomaly\cite{Aboubrahim:2021ohe}. 
This letter is a brief discussion of the main results of the consequences of 
hidden sector initial conditions at the reheat temperature on thermal relics
and a  more detailed version of the analysis will appear in \cite{Li:2023nez}.  In the following 
 we discuss some of the observable consequences of a hot vs a cold hidden 
 sector at the reheat time.
 
    \section{Thermal effects on observables}  
 \subsection{Dark matter relics}
  As the preceding discussion indicates the visible and the hidden sectors will
  in general be in different heat baths. In the presence of couplings between the two sectors even feeble, a consistent analysis requires that one carry out a synchronous thermal evolution of the two sectors. Such a synchronous
  evolution requires solution to  $\xi(T)$ given by  Eq.(\ref{xi1}).
   A solution to $\xi(T)$ also requires a simultaneous solution to 
   the yield equations for the dark fermion $D$ and the dark photon $\gamma'$
   which results from the $U(1)_X$ gauge field acquiring mass. 
  We exhibit below the yield equations 
  \bea
\label{yield1}
&\frac{dY_D}{dT}= F(T)\Big[\left<\sigma v\right>_{D\bar{D} \rightarrow i\bar{i} }(T)Y_D^{eq}(T)^2\nonumber\\
   & -\left<\sigma v\right>_{D\bar{D} \rightarrow \gamma'\bar{\gamma'} }(T_h)Y_D(T_h)^2
     +\left<\sigma v\right>_{\gamma'\bar{\gamma'} \rightarrow D\bar{D} }(T_h)Y_{\gamma'}(T_h)^2 \Big]\,,\label{yeald1} 
   \eea
     \bea
  \label{yield2}
 &  \frac{dY_\gamma'}{dT} = F(T) \Big[\left<\sigma v\right>_{D\bar{D} \rightarrow \gamma'\bar{\gamma'} }(T_h)Y_D(T_h)^2\nonumber\\
 & -\left<\sigma v\right>_{\gamma'\bar{\gamma'} \rightarrow D\bar{D} }(T_h)Y_{\gamma'}(T_h)^2 \nonumber\\
 &   +\left<\sigma v\right>_{i\bar{i}\rightarrow \gamma' }(T)Y_{i}^{eq}(T)^2 -\left<\Gamma_{\gamma'\rightarrow i\bar{i}(T_h)}\right>Y_{\gamma'}(T_h) \Big]\,.\nonumber\\
 &  F(T)\equiv -\frac{\s}{H}\left( \frac{d\rho_v/dT}{{4\zeta\rho-4\zeta_h\rho_h}+j_h/H}\right)    
        \label{yield3}
\eea   
where $\s$ is the entropy density and $v$ the relative velocity.
Dark photon is unstable and decays via the process
$\gamma'\to 3 \gamma$ and the entire relic density arises from the dark Dirac
fermions $D$ and $\bar D$ so that 
\bea
    \Omega_D h^2 = {s_0m_D Y^0_D h^2}/{\rho_c}\,,
\eea
where $s_0$ is the current entropy density, $m_D$ is the mass of the D-fermion,
$Y^0_D$ is $Y_D$ 
at current times, and $h$ is the Hubble parameter $H_0$ today in units of 100km s$^{-1}$ Mpc$^{-1}$.  
Using the above set of equations one can carry out a synchronous evolution of
the visible and the hidden sectors and compute the ratio $\xi(T)$ (using the
visible sector as a clock) by solving the coupled set of equations involving
the $\xi(T)$ equation Eq.(\ref{xi1}) and the yield equations Eq.(\ref{yield1})
and (\ref{yield2}).\\

\begin{figure}[h]
   \centering
  \includegraphics[width=0.8\linewidth]{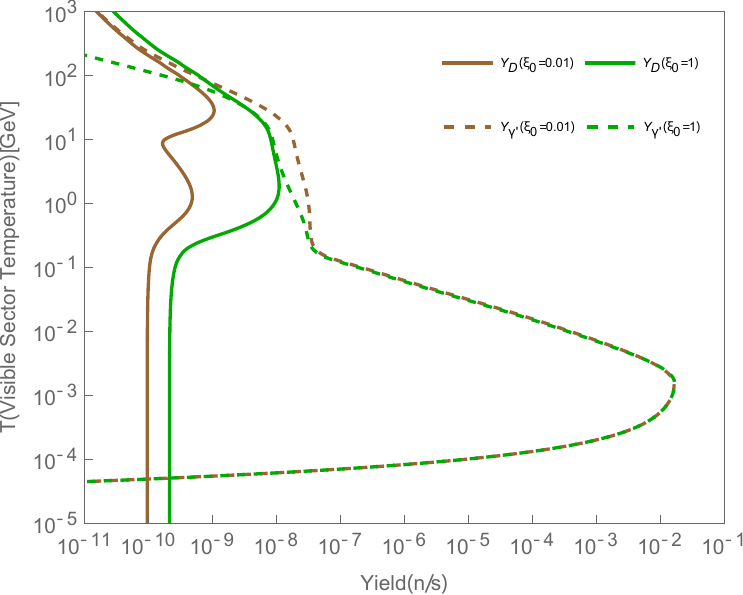}   
\caption{Yields of dark fermion (dark matter) and dark photon  for a cold hidden sector at $T_{RH}$,  i.e., 
$\xi_0=0.01$ (Brown), and a hot hidden sector at $T_{RH}$, i.e.,
$\xi_0=1$ (Green). The model parameters are $m_D = 2$GeV, $m_{\gamma'} = 1.22$MeV, $g_x = 0.019$, $\delta = 4\times10^{-9}$.The relic density for $\xi_0=0.01$ is 0.0524 while for $\xi_0=1$ is 0.117.
The shift in the relic density from an initially hot hidden sector to an initially
cold hidden sector is $\mathcal{O}(100\%)$.   }
\label{F1_zoom}
\end{figure}

  We note that the yield equations involve two different temperatures on the 
  right hand side in Eq.(\ref{yield1}) and Eq.(\ref{yield2}). Thus the objects
  \begin{align}
 \left<\sigma v\right>_{D\bar{D} \rightarrow \gamma'{\gamma'} }(T_h),
    \left<\sigma v\right>_{\gamma'{\gamma'} \rightarrow D\bar{D} }(T_h),
\left<\Gamma_{\gamma'\rightarrow i\bar{i}(T_h)}\right>,
\end{align}
appearing on the right hand side of Eq.(\ref{yield1}) and Eq.(\ref{yield2})
depend on the hidden sector temperature $T_h$ while the quantities 
\begin{align}
\left<\sigma v\right>_{i\bar{i} \rightarrow D\bar{D} }(T),
\left<\sigma v\right>_{i\bar{i}\rightarrow \gamma' }(T)
\end{align}
depend on the visible sector temperature $T$  indicating that a synchronous 
evolution of the thermal baths of the visible and the hidden sector is essential
for a consistent solution to $\xi(T), Y_D(T_h), Y_{\gamma'}(T_h)$. 
 However, here the initial conditions at the reheat time on the  hidden sector become relevant. Thus, as discussed earlier the two extreme possibilities
here are that at the reheat temperature the hidden sector either couples
to the inflaton as strongly as the visible sector does in which case 
$\xi_0\simeq 1$ and we have a hot hidden sector initially, or alternately it does not couple to the inflaton at all or couples very feebly in which case $\xi_0\sim 0$
in which case we have a cold hidden sector initially.
We exhibit the effects of the initial 
conditions on the hidden sector in Fig.(\ref{F1_zoom})
The analysis shows that the $\xi_0=1$ initial condition (hot hidden sector) gives a larger yield by $O(100\%)$ or more than the $\xi_0=0.01$ initial condition (cold hidden sector) highlighting the significant effect that the hidden initial condition has
on the yield and on the relic density. 
\subsection{Sommerfeld enhancement of dark matter cross sections}
We discuss now the effects on Sommerfeld enhancement of dark matter cross sections
 when the hidden sector is hot vs cold at the reheat temperature in the early
 universe.  The dark matter cross sections arise from various contributions,
 i.e., $DD\to DD$, $\bar D\bar D \to \bar D \bar D$, and $D\bar D\to D\bar D$.
 The interactions governing the scatterings  arise from the exchange of dark photons and in  the non-relativistic limit the potential governing the scattering takes the form 
\begin{align}
V(r)=\pm \frac{(g_x)^2}{4\pi }\frac{e^{-m_{\gamma'}r}}{r}\,.
\label{potential}
\end{align}
Here $DD\to DD$ and $\bar D\bar D \to \bar D \bar D$ scattering yield a 
(repulsive) Yukawa potential with a plus sign while the $D\bar D\to D\bar D$ scattering yields (an attractive) Yukawa potential with a negative sign. However, at low velocities non-perturbative effects
via exchange of multiple dark photons become significant and must be
taken into account. These effects are typically summarized by the Sommerfeld
enhancement factor $S_E$ so that for the scattering process $A+B\to A+B$ 
one writes
\bea
 (\sigma_{AB} v)= S_E ({\sigma^0_{AB}} v)\,.
\eea
where $({\sigma^0_{AB}} v)$ is the Born approximation and $v$ is the
relative velocity of the colliding particles. 
Such non-perturbative effects generated by the repeated exchange of a
dark photon or from the exchange of some other mediator has been 
discussed by a number of previous authors
  (see, e.g.,\cite{Lattanzi:2008qa,Arkani-Hamed:2008hhe,Cassel:2009wt,Cirelli:2007xd,Bringmann:2016din,Feng:2009mn} and the references therein). \\
    
\begin{figure}[h]
\begin{center}
  \includegraphics[width=0.4\linewidth]{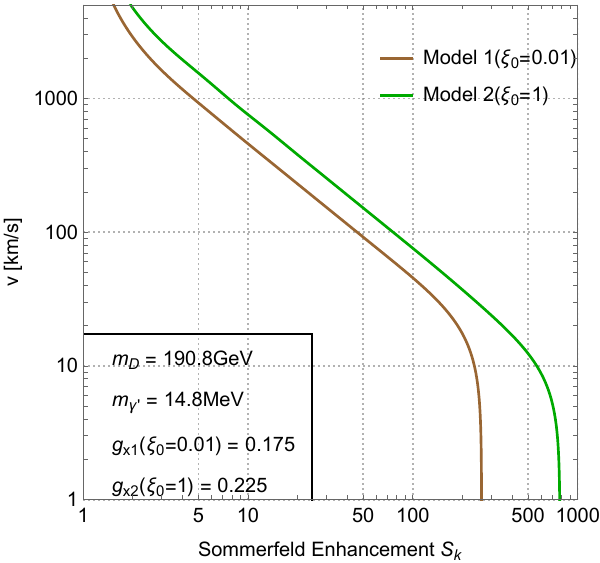} 
  \includegraphics[width=0.4\linewidth]{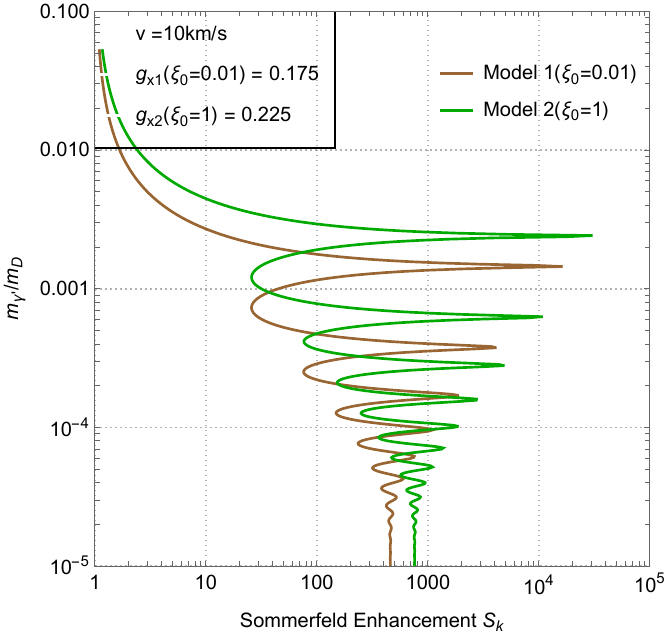}   
    \end{center}
\caption{ 
An exhibition of the effect of a hot vs a cold hidden sector at reheat on the
S-wave Sommerfeld enhancement of dark matter cross section 
for an attractive Yukawa potential.
The model parameters are $m_D = 190.8$GeV, $m_{\gamma'} = 14.8$MeV, $\delta = 35.8\times10^{-9}$. Left panel: Sommerfeld enhancement factor at different relative velocities for a hot dark sector ($\xi_0 =1$) and a cold dark sector ($\xi_0=0.01$).  To keep the relic density $\sim 0.12$, we choose $g_{x1} = 0.175$ for $\xi_0 = 0.01$ (Brown) and $g_{x2}=0.225$ for $\xi_0=1$ (Green). Right panel: Sommerfeld enhancement factor v.s. $m_D/m_{\gamma'}$ with $g_{x1}$ and $g_{x2}$ from left panel.}
\label{fig:Sommerfeld}
\end{figure}

 To take account of non-perturbative effects we  numerically solve the 
radial Schrödinger equation given by
\bea
p^2R_{l}+ \frac{d^2R_{l}}{dr^2}+\frac{2}{r}\frac{dR_{l}}{dr}-\frac{l(l+1)R_{l}}{r^2}
 -2\mu V(r)R_{l}=0,
\label{Eq:SE}
\eea
where $p$ is the particle momentum, $\mu$ is the reduced mass 
 and $V(r)$ is the Yukawa potential given by Eq.(\ref{potential}). 
Defining $x = p r$ and  $R_{p,l} = Npu_l(x)/x$ leads to following equation for
$u_l(x)$\cite{Iengo:2009ni}
\bea
&\left(\frac{d^2}{dx^2}+1-\frac{l(l+1)}{x^2}- \frac{2ae^{-bx}}{x}\right)u_l(x)=0,\nonumber\\
&a=\pm \frac{\mu g_X^2}{4 \pi p}, ~~b= \frac{m_{\gamma'}}{p}.
\label{Req}
\eea 
 The differential equation Eq.(\ref{Req})  has a solution of the form:
\bea
    \Phi_l(x)_{x\rightarrow\infty} \rightarrow C\sin(x-\frac{l\pi}{2}+\delta_l),
    \label{eq:DEsol}
\eea
where $\delta_l$ is the $l-$th partial wave phase shift.
The Sommerfeld enhancement for the $l$-th partial wave cross-section 
for the case of the Yukawa potential in then given by 
\bea
    \sigma_{l}={S_{E}}_l\cdot\sigma_{0,l}\,,
\eea
where \cite{Iengo:2009ni},
    ${S_{E}}_l=({1\cdot 3\cdots(2l+1)}/{C})^2$.
Using Eq. (\ref{eq:DEsol}), we get
\bea
C^2 &= \Phi^2_l(x)_{x\rightarrow\infty}+ \Phi^2_l(x-\frac{\pi}{2})_{x\rightarrow\infty}\,,\nonumber\\
{S_{E}}_l &= \frac{((2l+1)!!)^2}{\Phi^2_l(x)_{x\rightarrow\infty}+\Phi^2_l(x-\frac{\pi}{2})_{x\rightarrow\infty}}\,.
\eea
The analysis gives an enhancement of dark matter cross section at low collision velocities for attractive potentials and a suppression for the case of repulsive potential. The analysis shows that the enhancement is very sensitive to  $\xi_0$.
 In Fig.(\ref{fig:Sommerfeld})
  we exhibit this sensitivity. Here  one
finds that an initially hot hidden sector (i.e., $\xi_0=1$) gives a Sommerfeld
enhancement which could be
order few times larger relative to the case for an initially cold hidden sector.
\subsection{$N_{\rm eff}$ at BBN for hot vs cold hidden sector}
\begin{figure}[h]
   \centering   
           \includegraphics[width=0.5\linewidth]{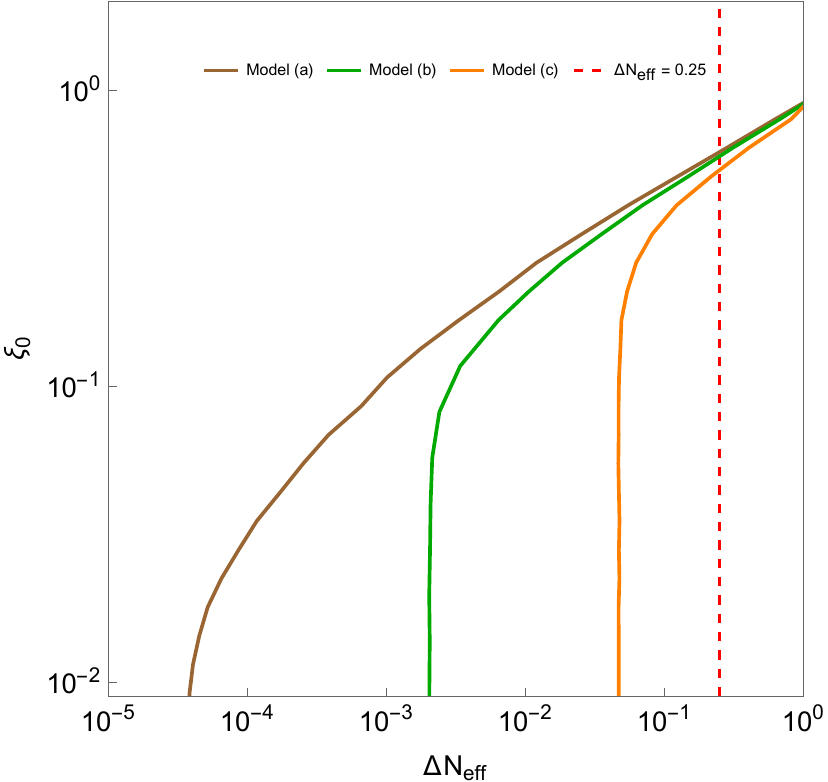}      
\caption{
Exhibition of the dependence of $\Delta N_{\rm eff}$ at BBN
   time on $\xi_0$ in the range $\xi_0=(0,1)$. Models (a)-(c) are defined by the
   parameter set $\{m_D,m_{\gamma'},g_X,\delta\}$ with value sets: (a):
   $\{0.767$GeV, $0.406$MeV, $0.00984$, $,2.88\times10^{-9}\}$; (b):
   $\{0.548$GeV, $,0.618$MeV, $ 0.0121$, $87.0\times10^{-9}\}$; (c):
  $\{0.796$GeV, $0.960$MeV, $0.0159$, $654\times10^{-9}\}$. The analysis shows that $\Delta N_{\rm eff}$ at BBN  can vary between $\Delta N_{\rm eff}=1$ for a hot hidden sector $(\xi_0=1)$  at the
   reheat and $\Delta N_{\rm eff}=10^{-5}$ for a cold hidden sector $(\xi_0=0)$ 
   at the reheat due to the suppression factor $(T_h/T)^4$ pointing to the 
  precision needed in the computation of $\xi(T_{\text{BBN}})$. The dashed line
  indicates the approximate upper limit of the error corridor for new degrees of freedom in model building. The analysis is consistent
with all known constraints on the hidden sector~\cite{Aboubrahim:2022qln} }
\label{fig-a}
\end{figure}

$N_{\rm eff}$ represents the number of massless neutrino degrees of freedom
beyond those of the standard model and is constrained by experimental 
data on the possible corridor between experiment and the standard model 
prediction in which it can reside. It acts as a strong constraint on model
building which involves new degrees of freedom that contribute to 
$\Delta N_{\rm eff}$. Thus let us suppose that the hidden sector has 
$g^h_{\rm eff}(T_h)$ massless
degrees of freedom at temperature $T_h$ which is synchronous with temperature
$T$ in the visible sector. In this case its contribution to $\Delta N_{\rm eff}$ is given by
\begin{align}
\Delta N_{\rm eff}&=\frac{4}{7} g^h_{\rm eff}(T_h)\left(\frac{11}{4}\right)^{4/3}
\left(\frac{T_h}{T}\right)^4\,,
\label{neff}
\end{align}
The standard model prediction for $N_{\rm eff}$ is  3.06 while the combined 
result from the Planck Collaboration~\cite{Planck:2018vyg}  and the joint 
BBN analysis of deuterium/helium abundance gives 
$N_{\rm eff}^{\rm exp}= 3.41\pm 0.45$. A conservative constraint on the extra
degrees of freedom is $\Delta N_{\rm eff}\leq 0.25$. We may contrast this
with the dispersion in $\Delta N_{\rm eff}$ created by the choice of a hot 
initial hidden sector or a cold initial hidden sector as illustrated in 
Fig(\ref{fig-a}) for three value sets for the parameters $\{m_D,m_{\gamma'},g_x,\delta\}$.
This figure illustrates a huge effect arising from the initial
conditions for the hidden sector due to the factor
$g^h_{\rm eff}(T_h)(T_h/T)^4$ which calls for an accurate computation of $\xi(T)$ for a reliable estimate of $\Delta N_{\rm eff}$ for a hidden sector model. \\
A comment is in order regarding Eq.(\ref{neff}) which requires that 
the hidden sector be in thermal equilibrium. This comes about as follows:
while the hidden sector is not in thermal equilibrium with the visible sector 
because of feeble couplings between them, this does not apply to internal
thermal equilibrium for the hidden sector. This is so because the couplings between the dark photons and  the dark fermions and among other dark
particles that may be around are not feeble but normal strength and thermal
equilibrium is established fairly quickly in thermal evolution. Further, 
$g^h_{eff}(T_h)$ is temperature dependent thus the temperature dependence
for the hidden sector degrees is not exactly $T_h^4$ but governed the $T_h$
dependence arising from the product $g^h_{eff}(T_h) T^4_h$. The exact computation of
$g^h_{\text{eff}}$ is done via thermal integrals and is exhibited in Appendix A.
 A further discussion of this topic can be found in ~\cite{Aboubrahim:2020lnr,Aboubrahim:2021ycj,Aboubrahim:2022bzk,Li:2023nez,
 Aboubrahim:2022gjb}.

\subsection{On the validity of separate entropy conservation approximation 
of co-moving visible and hidden sector volumes\label{ent1}}
 In the thermal evolution of the visible and the hidden sector from early times
  to later times a decoupling approximation is often used which assumes 
  that the entropy densities of the visible and the hidden sectors are 
  separately conserved in co-moving volumes.  This leads to the
  result that the ratio
  $s_h(T)/s_v(T)$ remains unchanged as the temperatures evolves from 
  the reheat temperature $T_0=T_{RH}$ down to the  temperature at   
   BBN time and to the current temperature. This assumption gives the 
relation    
   \begin{align} 
\frac{h^h_{eff}(\xi(T)T)}{h^v_{eff}(T)}\xi^3(T)
=\frac{h^h_{eff}(\xi(T_0)T_0)}{h^v_{eff}(T_0)}\xi^3(T_0)
\label{entropy-con}
\end{align}
where we used $T_{h}=\xi(T) T$ and $T_{0h}= \xi_0 T_0$.  
Eq.(\ref{entropy-con}) allows a computation of $\xi(T)$ using degrees of freedom at different temperatures. However, one may note that Eq.(\ref{entropy-con}) 
 has a highly non-linear dependence on $\xi(T)$ and one needs a numerical integration using thermal integrals. Here for the hidden sector we will use the thermal integrals for the entropy
degrees of freedom for $\gamma'$ and $D$  as given below~\cite{Hindmarsh:2005ix,Husdal:2016haj}
\begin{align}
h^{\gamma'}_{\rm eff}(T_h)=\frac{45}{4\pi^4}\int^{\infty}_{x_{h\gamma'}}
\frac{\sqrt{x^2-x_{h\gamma'}^2}}{e^x-1}(4x^2-x_{h\gamma'}^2)dx,
\end{align}
\begin{align}
  h^{D}_{\rm eff}(T_h)=\frac{15}{\pi^4}\int^{\infty}_{x_{hD}}\frac{\sqrt{x^2-x_{hD}^2}}{e^x+1}(4x^2-x_{hD}^2)dx\,,
\end{align}
where $x_{h\gamma'}=m_{\gamma'}/(T_h)= m_{\gamma'}/(\xi(T) T)$
and $x_{hD}= m_{D}/(\xi(T) T)$. For the visible sector thermal integrals
of the above type are not known because of hadronisation of quarks and 
gluons and the degrees of freedom are given in terms of a table or  a curve
as a function of temperature~\cite{Hindmarsh:2005ix,Husdal:2016haj}. \\

\begin{figure}[h]
    \centering
   \includegraphics[width=0.45\linewidth]{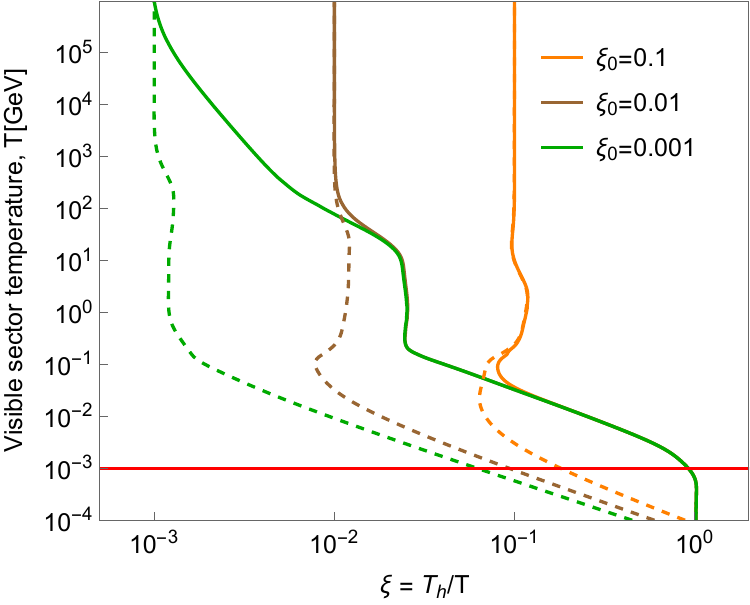}
       \includegraphics[width=0.45\linewidth]{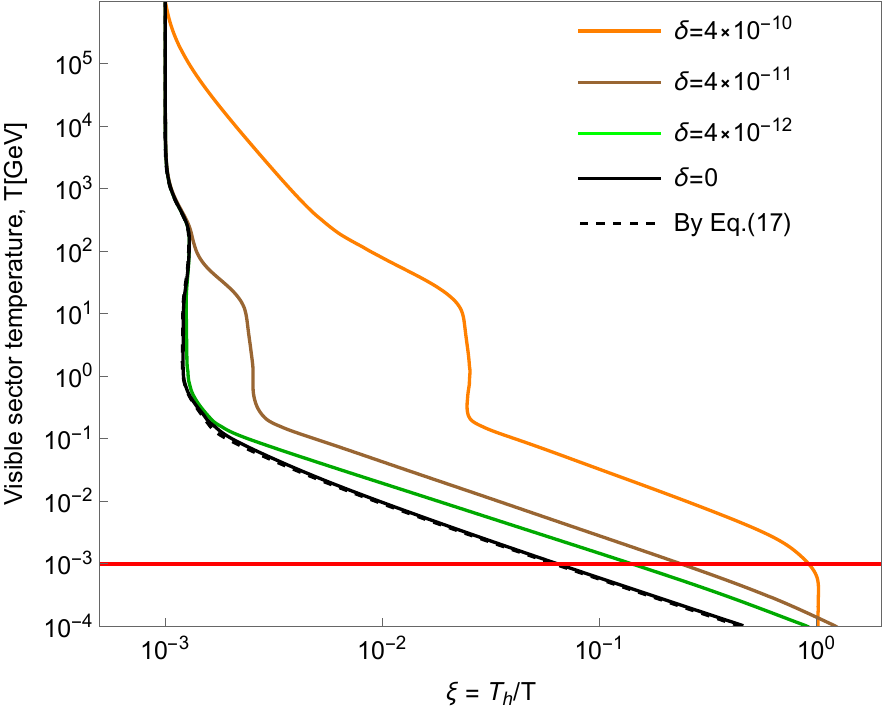}
     \caption{
    Evolution of $\xi(T)$ with different initial condition using Eq.(\ref{xi1}) of this paper (solid) and using the approximation of entropy conservation (dashed). Left panel: Here
$\delta = 4\times 10^{-10}$ and analysis is given for three 
widely different values of $\xi_0$, i.e., $\xi_0=0.001,\xi_0=0.01,
  \xi_0=0.1$.
 Right panel:  Here $\xi_0=0.001$ and an analysis for several different values 
 for $\delta$ in the range $\delta=0$ to $\delta=10^{-10}$ is
 exhibited. The rest of parameters are chosen so that $m_D = 2$ GeV, $m_{\gamma'} = 1.22$ MeV, $g_X = 0.019$.}
    \label{fig:entropy-approx}
\end{figure}

 Fig.~\ref{fig:entropy-approx} gives a comparison of the evolution of $\xi(T)$
 as a function of the temperature $T$ of the visible sector using the exact formula
 of Eq.(\ref{xi1}) (solid lines) vs the one using the approximation of entropy conservation of the visible and the hidden sector separately in comoving volumes  given  by dashed lines. Thus the left panel gives the analysis for different values of $\xi_0$. Here one finds  significant deviations of the 
 approximate results from the exact ones with the worst case occurring for the
 smallest $\xi_0$ case corresponding to the coldest hidden sector at the reheat
 temperature.
 The right panel gives the result for different values of the kinetic mixing parameter
 $\delta$ for a fixed value of $\xi_0$. Here one finds that even for very feeble couplings with 
 $\delta$ as small as $\delta=10^{-12}$ there are significant deviations of the predictions on $\xi(T)$ at BBN time between the exact and the approximate. 
 Thus, our conclusion, is that entropy conservation approximation separately for co-moving sectors of the visible and hidden sectors in thermal evolution is not suitable in general for precision cosmology.
\section{Conclusion}
The analysis discussed here exhibits the fact that the thermal condition of
the hidden sector at reheat temperature affects observables related to 
thermal relics. Thus assumptions of a hot vs a cold hidden sector at reheat can
lead   up to $O(100\%)$ shift on predicted values of observables
and for $\Delta N_{\text{eff}}$ by as much as a factor of $10^5$ due to the 
large variation generated by the factor $(\frac{T_h}{T})^4$ as $T_h/T$ varies. 
 It is also shown that the approximation of using entropy conservation in comoving volumes for the visible and the hidden sectors is invalid even for very feeble couplings between the visible and the hidden sectors.

\section*{Acknowledgements}
This research was supported in part by the NSF Grant PHY-2209903.\\

\noindent
\section{Appendix A: Energy density of hidden sector}
Assuming for illustration just dark photon ($\gamma'$) and dark fermion ($D$) in the hidden sector, the energy density of the hidden 
sector $\rho_h(T_h)$ is given by 
\begin{align}
  \rho_h (T_h)&=  \rho_{\gamma'}(T_h)+\rho_{D}(T_h) = \frac{\pi^2}{30}g_{eff}^{h}(T_h)T_h^4,
  \nonumber\\
    \rho_{\gamma'}(T_h) &= \frac{\pi^2}{30}g_{eff}^{\gamma'}(T_h)T_h^4\,,
    ~~\rho_{D}(T_h) = \frac{\pi^2}{30}g_{eff}^{D}(T_h)T_h^4\,  
    \label{density2}
\end{align}
\begin{align}
 g^h_{\rm eff}(T_h) = &g^{\gamma'}_{\rm eff}(T_h)+g^D_{\rm eff}(T_h)\nonumber\\
 =&\frac{45}{\pi^4}\int^{\infty}_{x_\gamma'}\frac{\sqrt{x^2-x_{\gamma'}^2}}{e^x-1}x^2dx +\frac{60}{\pi^4}\int^{\infty}_{x_D}\frac{\sqrt{x^2-x_{D}^2}}{e^x+1}x^2dx\,. 
\label{geff-heff}
\end{align}
where $x_{\gamma'}= m_{\gamma'}/T_h$, and 
$x_{D}= m_{D}/T_h$. Thus $g_{\rm eff}^h(T_h)$ is temperature dependent 
and the effective temperature that enters in Eq.(\ref{neff}) is not just $T_h^4$
but $g^h_{\rm eff}(T_h)T_h^4$.

\bibliographystyle{unsrt}

\begin{thebibliography}{99}
\bibitem{Nath:2016qzm}
P.~Nath,
Cambridge University Press, 2016,
ISBN 978-0-521-19702-1, 978-1-316-98396-6
doi:10.1017/9781139048118

\bibitem{Candelas:1985en}
P.~Candelas, G.~T.~Horowitz, A.~Strominger and E.~Witten,
Nucl. Phys. B \textbf{258}, 46-74 (1985)
doi:10.1016/0550-3213(85)90602-9

\bibitem{Polchinski:1996na}
J.~Polchinski,
[arXiv:hep-th/9611050 [hep-th]].

\bibitem{Hill:2000mu}
C.~T.~Hill, S.~Pokorski and J.~Wang,
Phys. Rev. D \textbf{64}, 105005 (2001)
doi:10.1103/PhysRevD.64.105005
[arXiv:hep-th/0104035 [hep-th]].

\bibitem{Aboubrahim:2020lnr}
A.~Aboubrahim, W.~Z.~Feng, P.~Nath and Z.~Y.~Wang,
Phys. Rev. D \textbf{103}, no.7, 075014 (2021)
doi:10.1103/PhysRevD.103.075014
[arXiv:2008.00529 [hep-ph]].

\bibitem{Aboubrahim:2021ycj}
A.~Aboubrahim, W.~Z.~Feng, P.~Nath and Z.~Y.~Wang,
JHEP \textbf{06}, 086 (2021)
doi:10.1007/JHEP06(2021)086
[arXiv:2103.15769 [hep-ph]].

\bibitem{Aboubrahim:2022bzk}
A.~Aboubrahim and P.~Nath,
JHEP \textbf{09}, 084 (2022)
doi:10.1007/JHEP09(2022)084
[arXiv:2205.07316 [hep-ph]].

\bibitem{Feng:2008mu}
J.~L.~Feng, H.~Tu and H.~B.~Yu,
JCAP \textbf{10}, 043 (2008)
doi:10.1088/1475-7516/2008/10/043
[arXiv:0808.2318 [hep-ph]].

\bibitem{Ertas:2021xeh}
F.~Ertas, F.~Kahlhoefer and C.~Tasillo,
JCAP \textbf{02}, no.02, 014 (2022)
doi:10.1088/1475-7516/2022/02/014
[arXiv:2109.06208 [astro-ph.CO]].

\bibitem{Holdom:1985ag}
B.~Holdom,
Phys. Lett. B \textbf{166}, 196-198 (1986)
doi:10.1016/0370-2693(86)91377-8

\bibitem{Kors:2004dx}
B.~Kors and P.~Nath,
Phys. Lett. B \textbf{586}, 366-372 (2004)
doi:10.1016/j.physletb.2004.02.051
[arXiv:hep-ph/0402047 [hep-ph]].

\bibitem{Feldman:2007wj}
D.~Feldman, Z.~Liu and P.~Nath,
Phys. Rev. D \textbf{75}, 115001 (2007)
doi:10.1103/PhysRevD.75.115001
[arXiv:hep-ph/0702123 [hep-ph]].

\bibitem{Du:2022fqv}
M.~Du, Z.~Liu and P.~Nath,
Phys. Lett. B \textbf{834}, 137454 (2022)
doi:10.1016/j.physletb.2022.137454
[arXiv:2204.09024 [hep-ph]].

\bibitem{Patt:2006fw}
B.~Patt and F.~Wilczek,
[arXiv:hep-ph/0605188 [hep-ph]].

\bibitem{Cheung:2007ut}
K.~Cheung and T.~C.~Yuan,
JHEP \textbf{03}, 120 (2007)
doi:10.1088/1126-6708/2007/03/120
[arXiv:hep-ph/0701107 [hep-ph]].

\bibitem{Aboubrahim:2021ohe}
A.~Aboubrahim, P.~Nath and Z.~Y.~Wang,
JHEP \textbf{12}, 148 (2021)
doi:10.1007/JHEP12(2021)148
[arXiv:2108.05819 [hep-ph]].

\bibitem{Li:2023nez}
J.~Li and P.~Nath,
Phys. Rev. D \textbf{108}, no.11, 115008 (2023)
doi:10.1103/PhysRevD.108.115008
[arXiv:2304.08454 [hep-ph]].

\bibitem{Lattanzi:2008qa}
M.~Lattanzi and J.~I.~Silk,
Phys. Rev. D \textbf{79}, 083523 (2009)
doi:10.1103/PhysRevD.79.083523
[arXiv:0812.0360 [astro-ph]].

\bibitem{Arkani-Hamed:2008hhe}
N.~Arkani-Hamed, D.~P.~Finkbeiner, T.~R.~Slatyer and N.~Weiner,
Phys. Rev. D \textbf{79}, 015014 (2009)
doi:10.1103/PhysRevD.79.015014
[arXiv:0810.0713 [hep-ph]].

\bibitem{Cassel:2009wt}
S.~Cassel,
J. Phys. G \textbf{37}, 105009 (2010)
doi:10.1088/0954-3899/37/10/105009
[arXiv:0903.5307 [hep-ph]].

\bibitem{Cirelli:2007xd}
M.~Cirelli, A.~Strumia and M.~Tamburini,
Nucl. Phys. B \textbf{787}, 152-175 (2007)
doi:10.1016/j.nuclphysb.2007.07.023
[arXiv:0706.4071 [hep-ph]].

\bibitem{Bringmann:2016din}
T.~Bringmann, F.~Kahlhoefer, K.~Schmidt-Hoberg and P.~Walia,
Phys. Rev. Lett. \textbf{118}, no.14, 141802 (2017)
doi:10.1103/PhysRevLett.118.141802
[arXiv:1612.00845 [hep-ph]].

\bibitem{Feng:2009mn}
J.~L.~Feng, M.~Kaplinghat, H.~Tu and H.~B.~Yu,
JCAP \textbf{07}, 004 (2009)
doi:10.1088/1475-7516/2009/07/004
[arXiv:0905.3039 [hep-ph]].

\bibitem{Iengo:2009ni}
R.~Iengo,
JHEP \textbf{05}, 024 (2009)
doi:10.1088/1126-6708/2009/05/024
[arXiv:0902.0688 [hep-ph]].

\bibitem{Planck:2018vyg}
N.~Aghanim \textit{et al.} [Planck],
Astron. Astrophys. \textbf{641}, A6 (2020)
[erratum: Astron. Astrophys. \textbf{652}, C4 (2021)]
doi:10.1051/0004-6361/201833910
[arXiv:1807.06209 [astro-ph.CO]].

\bibitem{Aboubrahim:2022gjb}
A.~Aboubrahim, M.~Klasen and P.~Nath,
JCAP \textbf{04}, no.04, 042 (2022)
doi:10.1088/1475-7516/2022/04/042
[arXiv:2202.04453 [astro-ph.CO]].

\bibitem{Aboubrahim:2022qln}
A.~Aboubrahim, M.~M.~Altakach, M.~Klasen, P.~Nath and Z.~Y.~Wang,
JHEP \textbf{03}, 182 (2023)
doi:10.1007/JHEP03(2023)182
[arXiv:2212.01268 [hep-ph]].

\bibitem{Hindmarsh:2005ix}
M.~Hindmarsh and O.~Philipsen,
Phys. Rev. D \textbf{71}, 087302 (2005)
doi:10.1103/PhysRevD.71.087302
[arXiv:hep-ph/0501232 [hep-ph]].

\bibitem{Husdal:2016haj}
L.~Husdal,
Galaxies \textbf{4}, no.4, 78 (2016)
doi:10.3390/galaxies4040078
[arXiv:1609.04979 [astro-ph.CO]].

\end{thebibliography}

\end{document}